\documentclass[prb,twocolumn,showpacs,nofootinbib]{revtex4}
\usepackage[dvips]{graphicx}
\usepackage{color,array,dcolumn}
\usepackage{amsmath}
\usepackage{amssymb}

\begin{document}

\title{Polarization of a quasi two-dimensional repulsive Fermi gas with Rashba spin-orbit coupling: a variational study}
\author{A. Ambrosetti$^{1}$}
\email{alberto.ambrosetti@unipd.it}
\author{G. Lombardi$^{2}$, L. Salasnich$^{1}$, P.L. Silvestrelli$^{1}$}
\author{F.Toigo$^{1}$}
\affiliation{1) Dipartimento di Fisica, University of Padova, via Marzolo 8, I--35131, Padova, Italy.}
\affiliation{2) TQC, Universiteit Antwerpen, Universiteitsplein 1, B-2610 Antwerpen, Belgium}

\begin{abstract}
\date{\today}

Motivated by the remarkable experimental  control of
synthetic gauge fields in  ultracold atomic systems,
we investigate the effect of an artificial Rashba spin-orbit coupling
on the spin polarization of a two-dimensional repulsive Fermi gas.
By using a variational many-body wavefunction, based on a suitable spinorial structure,
we find that the polarization properties of the system are indeed controlled by
the interplay between spin-orbit coupling and repulsive interaction.
In particular, two main effects are found:
1) The Rashba coupling determines a gradual increase of the degree of
 polarization beyond the critical repulsive interaction strength, at variance with conventional
2D Stoner instability. 2) The critical interaction strength,
above which finite polarization is developed, shows a dependence on the Rashba coupling,
{\it i.e.} it is enhanced in case the Rashba coupling exceeds a critical value.
A simple analytic expression for the critical interaction strength is further derived
in the context of our variational formulation, which allows for a straightforward
and insightful analysis of the present problem.

\end{abstract}

\maketitle

\section{Introduction}
The Rashba spin-orbit (SO) coupling\cite{Rashba}, initially considered in semiconductor devices such as quantum wells\cite{Kouwenhoven,Steward,nitta,engels,espab,nittathick},
 has been recently reproduced in ultracold atoms \cite{Dalibard, cold,so2,so3} by means of externally applied controlled laser beams. 
The relevance of this  achievement can be understood in relation to the extremely high degree of control achievable in these systems.
In fact, the absence of interfering phonons present in solid state aggregates 
and the precise tunability of external laser sources, make ultracold atoms invaluable candidates for the experimental investigation of
the most delicate aspects of quantum mechanics, such as SO effects.

Following this success, a large number of theoretical studies recently focused on the intricate effects of Rashba
or Dresselhaus\cite{Dresselhaus} SO couplings in Bose-Einstein condensates\cite{bc1,bc2,bc3,bc4,bc5,bc6,bc7,bc8,bc9},  in superfluid
Fermions at the BCS-BEC crossover\cite{bec1,bec2,bec3,bec4,bec5,bec6,bec7,bec8,bec9,bec10,bec11,bec12,bec13,bec14,bec15,bec16,bec17,bec18,bec19,bec21},
 and in the quasi-ideal Fermi gas\cite{sal1}.
Yet, little is known so far  concerning the possible effects of SO couplings in the repulsive atomic
Fermi gas, a system which is widely known for its intrinsic and peculiar polarization properties\cite{pilati,fratini,macdonald,cond2d,cond3d,condprl}.

In this manuscript we will consider a two-component bi-dimensional (2D) atomic Fermi assembly in presence of Rashba SO coupling. 
Low dimensional atomic Fermi systems are of particular relevance in this context, in view of possible spintronics applications, 
and are currently undergoing experimental investigations\cite{exp2d1,exp2d2}, also in relation to their non-trivial spin transport properties\cite{expspin}.
In the present system, due to the non-local spin structure induced by the SO coupling, a conventional Hartree-Fock approach 
based on single-particle states whose spin is aligned along a fixed axis is not applicable.
We will thus introduce a suitable variational procedure, making use of an appropriate
functional form for the single particle orbitals, whose spin structure will depend on a variational
parameter.
This approach allows for a seamless connection between spin polarized and unpolarized regimes, through the minimization of
a unique energy function.

\section{Model}

The Hamiltonian of a uniform 2D repulsive Fermi gas in presence of Rashba spin-orbit coupling can be written as a sum of a one-body (1B) and a two-body (2B) contribution:
\begin{equation}
\hat{H}=\hat{H}_{1B}+\hat{H}_{2B}\,.
\end{equation}
In second quantization, the single particle contribution is defined as 
\begin{equation}
\hat{H}_{1B}=\int d^2 \mathbf{r} \, \hat{\Psi}^{\dagger}(\mathbf{r})\hat{h}_{sp}\hat{\Psi}(\mathbf{r})  \,,
\label{hsp}
\end{equation}
where the two-component quantum field operators
\begin{equation}
\hat{\Psi}(\mathbf{r})=  \left(
    \begin{array}{c}
      \hat{\psi_{\uparrow}}(\mathbf{r})  \\
      \hat{\psi_{\downarrow}}(\mathbf{r})
    \end{array} \right) 
\qquad  {\rm and} \qquad
\hat{\Psi}^{\dagger}(\mathbf{r})=\left(\hat{\psi}^{\dagger}_{\uparrow}(\mathbf{r}),\hat{\psi}^{\dagger}_{\downarrow}(\mathbf{r})  \right)
\end{equation}
are defined in terms of the operators $\hat{\psi}^{\dagger}_{\sigma}(\mathbf{r})$ and $\hat{\psi}_{\sigma}(\mathbf{r})$, which construct and annihilate one particle with spin $\sigma$ $(\sigma= \uparrow,\downarrow)$ at the position $\mathbf{r}=(x,y)$.
The single-particle operator $\hat{h}_{sp}$ in Eq. \eqref{hsp} accounts for both the kinetic energy and the Rashba potential, and is defined as
\begin{equation}
\hat{h}_{sp}=\frac{\hat{p}^2}{2m}\sigma_0+\lambda_R(\hat{p}_y\sigma_x-\hat{p}_x\sigma_y) \,,
\label{hsp2}
\end{equation}
where  $\hat{p}_{x,y}$ represents the momentum operator,  $\sigma_0$ is the 2D identity matrix, and $\sigma_{x,y}$ are the $x,y$ Pauli matrices.
The quantity $m$ represents the particle mass, while $\lambda_R$ is the Rashba coupling, accounting for the
tunability of the Rashba\cite{Rashba} SO coupling (the second term to the right in Eq.\eqref{hsp2}). Planck's constant $\hbar$ was set to 1 to simplify the notation.

As concerns the two-body contribution, 
at very low energy only the s-wave scattering will be relevant, and the interaction may be modelled - due to the Pauli principle - as a contact potential
 acting  only  between particles of opposite spins. In second quantization the repulsive two-body
interaction thus takes the  form:
\begin{equation}
\hat{H}_{2B}=\int d^2\mathbf{r} \,  g \,\hat{\psi}^{\dagger}_{\uparrow}(\mathbf{r}) \hat{\psi}^{\dagger}_{\downarrow}(\mathbf{r}) \hat{\psi}_{\downarrow}(\mathbf{r}) \hat{\psi}_{\uparrow}(\mathbf{r}) .
\end{equation}
The  coupling constant $g$ is positive, and it may be experimentally tuned by exploiting the Feshbach resonance mechanism\cite{feshbach,chin}.
Since an effectively 2D atomic gas can be realized by confining a three dimensional (3D) system along the $z$ direction
 within a small thickness $b$, a direct relation exists between the coupling constant $g$, the energetic and geometrical properties
of the system, and the 3D scattering length $a$, as discussed in Refs. (\onlinecite{bhaduri,petrov}).

Regarding the total density of particles, which will be fixed in our calculations, this is obtained as
the expectation value of the operator
\begin{equation}
\hat{n}=\frac{1}{L^2}\int d^2\mathbf{r} \, \left\{ \hat{\Psi}^{\dagger}_{\uparrow}(\mathbf{r}) \hat{\Psi}_{\uparrow}(\mathbf{r}) + \hat{\Psi}^{\dagger}_{\downarrow}(\mathbf{r}) \hat{\Psi}_{\downarrow}(\mathbf{r}) \right\}\,,
\end{equation}
where $L^2$ is the area of our 2D system.
Analogously, the spin polarization  is derived from the expectation value of the operator
\begin{equation}
\hat{m}=\frac{1}{L^2}\int d^2\mathbf{r} \, \left\{ \hat{\Psi}^{\dagger}_{\uparrow}(\mathbf{r}) \hat{\Psi}_{\uparrow}(\mathbf{r}) - \hat{\Psi}^{\dagger}_{\downarrow}(\mathbf{r}) \hat{\Psi}_{\downarrow}(\mathbf{r}) \right\} \,.
\end{equation}

\section{Variational Procedure}
In the absence of SO coupling, a straightforward HF (or mean field) approach could be applied to our system\cite{macdonald,cond2d}.
In fact, in this particular case the Hamiltonian commutes with the spin operators, and the single-particle
 wave functions can be safely chosen as eigenstates of the third Pauli matrix $\sigma_z$.
This choice leads to a clear distinction between spin up ($n_{\uparrow}$) ad spin down ($n_{\downarrow}$)
particle densities, and at any given value of $g$
the HF energy per volume can thus be expressed as a function of these quantities: $E(n_{\uparrow},n_{\downarrow})$.
A minimization of this function at fixed total density $n=n_{\uparrow}+n_{\downarrow}$, leads to 
zero polarization ($n_{\uparrow}=n_{\downarrow}$) for $g$ below the critical value $g_c=\pi/m$, and full polarization (either $n=n_{\uparrow}$ or $n=n_{\downarrow}$) at $g>g_c$ (Stoner instability\cite{stoner,pilati,fratini,macdonald,cond2d,cond3d,condprl}).

When including the Rashba SO coupling, however, the above procedure is not directly applicable. The main problem with a mean field description based
on $\sigma_z$ single-particle eigenstates is that the expectation value of the SO coupling on both spin-up and spin-down wave functions is identically zero. 
No SO energy contribution would thus be present in such a mean field approach.

A possible alternative, explored in recent studies of the 2D electron gas with Rashba SO coupling\cite{dmc,dmcdot,dmcs}, is to make use of single-particle eigenfunctions
of the Rashba coupling\cite{lipp}. Given the dependence of the SO coupling on momentum, these orbitals retain a plane wave structure, accompanied by a $k$-dependent spinorial form.
These wave functions provide information about the non-local spin structure induced by the SO coupling, leading in general to non-zero SO energy
contributions.
A straightforward use of Rashba eigenstates, however, is not possible in our case, since the expectation value of $\sigma_z$ on Rashba eigenstates is zero. 
A different approach must thus be tailored when aiming to study  the polarization properties of our system.

To combine the polarization properties of $\sigma_z$ eigenstates with the non-local structure induced by the Rashba coupling, we introduce a
variational many-body state $|\Phi^{\rm VAR}_{\zeta,n_+}\rangle$, depending on two variational parameters $\zeta$ and $n_+$
whose meaning will become clear in the following.

Physically, a favorable way of controlling the polarization of a given system is that of
applying an external magnetic field, inducing a spin-dependent coupling.
Following this idea, we define an auxiliary  Hamiltonian 
\begin{equation}
\hat{H}_{\zeta}=\int d^2 \mathbf{r} \, \hat{\Psi}^{\dagger}(\mathbf{r})\hat{h}_{\zeta}\hat{\Psi}(\mathbf{r})  \,,
\end{equation}
where $\hat{h}_{\zeta}=\hat{h}_{sp}+\hat{v}_{\zeta}$ and   $\hat{v}_{\zeta}=\zeta\sigma_z$.
This Hamiltonian only contains single-particle energy terms, and can favor, through the parameter
$\zeta$, the alignment or antialignment of the spin along the $z$ axis, while still retaining
the non-local Rashba SO coupling.
Hence, we explicitly construct our many-body variational wave function as an eigenstate of $\hat{H}_{\zeta}$:
\begin{equation}
\hat{H}_{\zeta}|\Phi^{\rm VAR}_{\zeta,n_+}\rangle =E^{(0)}_{\zeta,n_+} |\Phi^{\rm VAR}_{\zeta,n_+}\rangle
\label{e-egvl}
\end{equation}

The diagonalization of $\hat{H}_{\zeta}$ within the single-particle framework leads to the single-particle eigenenergies
\begin{equation}
\epsilon_{\pm}(\mathbf{k})=\frac{k^2}{2m}\pm\sqrt{\zeta^2+\lambda^2k^2}
\label{spen}
\end{equation}
corresponding to the single-particle eigenstates
\begin{equation}
\phi^{\pm}_{\mathbf{k}}(\mathbf{r})= c_{\pm,\mathbf{k}}
  \left(
    \begin{array}{c}
    \frac{\lambda(k_y+ik_x)}{\zeta \pm \sqrt{\zeta^2+\lambda^2k^2}}     \\
    1
  \end{array} \right) e ^{i\mathbf{k\cdot r}}
\label{rstates}
\end{equation}
where the normalization constants $c_{\pm,\mathbf{k}}$ are defined to be real and obey the equation:
\begin{equation}
c_{\pm,\mathbf{k}}=\left( \frac{\lambda^2 k^2}{(\zeta \pm \sqrt{\zeta^2+\lambda^2k^2})^2} +1 \right)^{-1/2}.
\end{equation}
Here $\mathbf{k}=(k_x,k_y)$ is a 2D wave vector.
As seen, two solutions exist for every fixed value of $\mathbf{k}$, corresponding to two distinct energy bands.
These will be hereafter labelled as $\pm$  for simplicity. 
We underline that these single particle states depend, through $\hat{v}_{\zeta}$ on the external parameter $\zeta$. This has not
been fixed yet, and will be used in the following as a variational parameter.

The variational many-particle wave function $|\Phi^{\rm VAR}_{\zeta,n_+}\rangle$
 is constructed as an antisymmetrized product of
the above single particle eigenstates, and is constrained to describe a fixed density of particles $n$
through the relation
\begin{equation}
\hat{n}\,|\Phi^{\rm VAR}_{\zeta,n_+}\rangle =n \,|\Phi^{\rm VAR}_{\zeta,n_+}\rangle.
\end{equation}
The total  density of particles is equal to the sum of the densities relative to the $+$ and $-$
 bands ($n=n_++n_-$). The densities of the $\pm$ bands ($n_{\pm}$) can in turn be 
 expressed in terms of the relative Fermi energies $\epsilon^F_{\pm}$, as
\begin{equation}
n_{\pm}=\frac{1}{(2\pi)^2}\int d^2\mathbf{k}\,\, \Theta(\epsilon^F_{\pm}-\epsilon_{\pm}(\mathbf{k}))\,,
\end{equation}
where $\Theta(x)$ is the Heaviside step function.
Given the constraint of fixed total density $n$, only $n_+$ can be regarded as an independent 
variable, controlling the relative $+/-$ band occupation in $|\Phi^{\rm VAR}_{\zeta,n_+}\rangle$.

Depending on the $\zeta$ value and the single-particle band occupation, the system will be allowed to develop both a 
 non-local Rashba spin texture, or a spin-polarized configuration.
The relevant spin configurations will thus be spanned by our theory, and the present approach will
represent an extension of the original variational procedure to a broader space of configurations.

Once the variational wave function has been defined, the energy is written as
\begin{equation}
E_{{\rm VAR}}(n_+,\zeta)=\langle\Phi^{\rm VAR}_{\zeta,n_+}|\hat{H}|\Phi^{\rm VAR}_{\zeta,n_+}\rangle \,.
\label{evar}
\end{equation}
This expression is a function of the parameter $\zeta$ and of the single particle band occupation
 $n_+$, and its minimization at fixed total density $n$  leads to the optimal
parameters $\bar{\zeta}$ and $\bar{n}_+$, from which the  wave function $|\Phi^{\rm VAR}_{\bar{\zeta},\bar{n}_+}\rangle$
is determined. We also stress that $E_{{\rm VAR}}(n_+,\zeta)$, differently from $E^{(0)}_{\zeta,n_+}$, accounts for the
presence of the two-body interaction.

Finally, after the optimal wave function $|\Phi^{\rm VAR}_{\bar{\zeta},\bar{n}_+}\rangle$ has been obtained through the variational procedure,
the polarization $P$ can be straightforwardly computed as
\begin{equation}
P(\bar{\zeta},\bar{n}_+)=\langle\Phi^{\rm VAR}_{\bar{\zeta},\bar{n}_+}|\hat{m}|\Phi^{\rm VAR}_{\bar{\zeta},\bar{n}_+}\rangle/ n
\label{P}
\end{equation}

As concerns the actual computation of $E_{{\rm VAR}}(n_+,\zeta)$, this can be simplified by making use
of the relation \eqref{e-egvl}. Details and analytical formulas regarding $E_{{\rm VAR}}(n_+,\zeta)$ and $P$
 are reported and discussed in detail in the Appendix.

\section{Results}
We present in this section the energy and polarization properties of the system at $T=0$, derived according to the variational
procedure introduced in  section III.
 In the following we will express lengths in units of $n^{-1/2}$, and energies in units of $\frac{n \hbar^2}{ m}$.

As already outlined in section II (see Appendix for details), if $\bar{\zeta}=0$ no polarization is present in the system,
since the single particle states on which $|\Phi^{\rm VAR}_{\zeta,n_+}\rangle$ relies coincide in this case with the Rashba
eigenstates.
On the other hand, if a polarized state is energetically favored with respect to a spin unpolarized configuration,
 a finite optimal $\bar{\zeta}$ will be found within the variational procedure.
\newline
\newline
\begin{figure}[ht]
\vspace{0.2cm}
\centering
\includegraphics[scale=0.115]{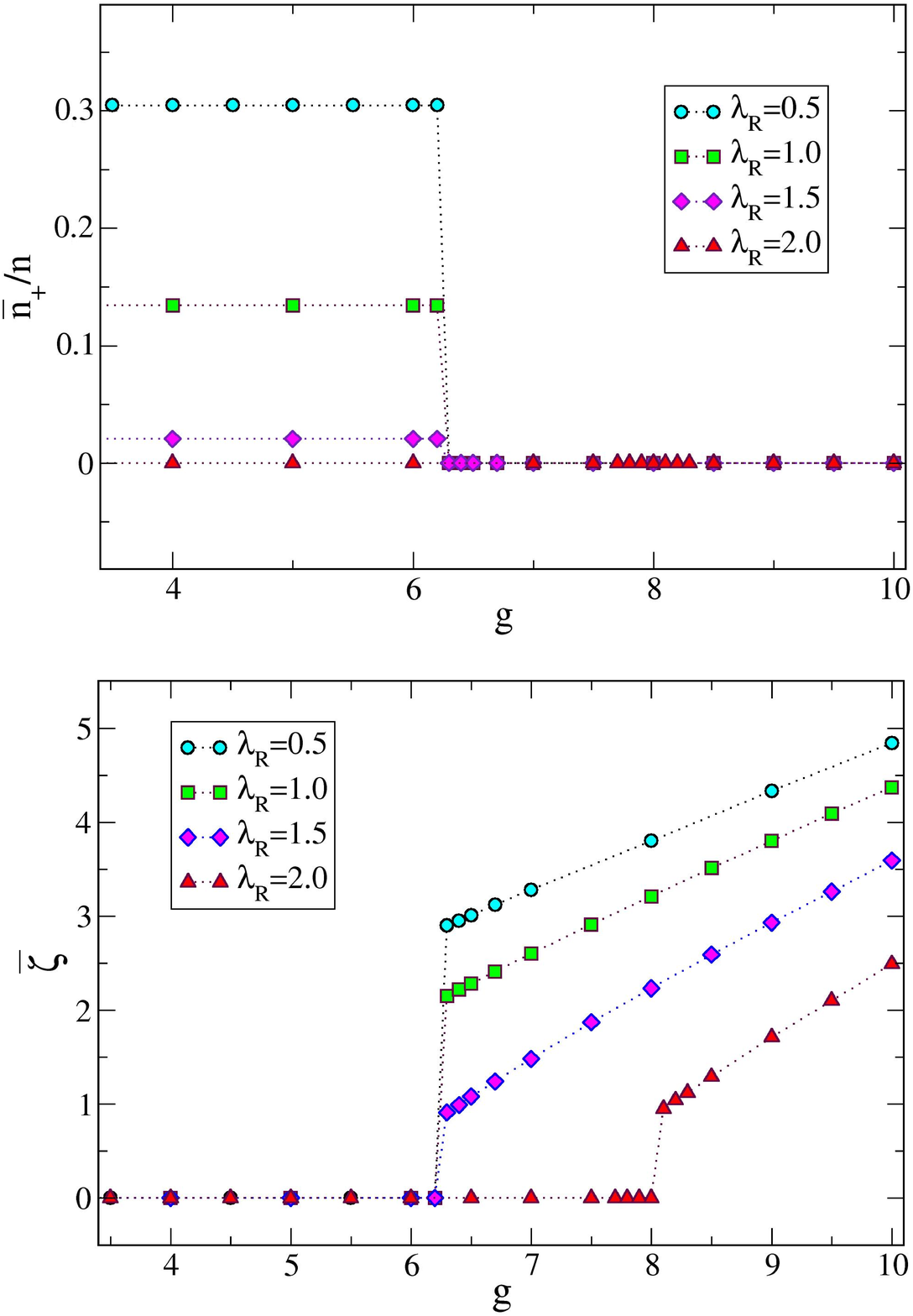}
\vspace{0.2cm}
\caption{(color online) Optimal variational parameters $\bar{n}_+$ and $\bar{\zeta}$, plotted in function of $g$ for different values of $\lambda_R$. 
 Units are specified in the text.}
\label{figura2}
\end{figure}

Clearly, depending on $g$ and $\lambda_R$, different optimal $\bar{n}_+$ and $\bar{\zeta}$ values will be found,
corresponding to different values of the polarization.

Interestingly, the introduction of a $\zeta$ dependent $|\Phi^{\rm VAR}_{\zeta,n_+}\rangle$ allows to consistently 
achieve lower variational energies with respect to using both Rashba states and $\sigma_z$ eigenvectors
in the polarized regime. According to the variational principle $|\Phi^{\rm VAR}_{\zeta,n_+}\rangle$ is thus expected
to provide a more accurate approximation to the polarized ground states.

To understand the behavior of $|\Phi^{\rm VAR}_{\zeta,n_+}\rangle$ in relation to the repulsive interaction
we analyze $\bar{n}_+$ and $\bar{\zeta}$ as a function of $g$ and $\lambda_R$  (see Fig. \ref{figura2}). Clearly,
$\bar{\zeta}$ is identically zero below $g_c$, while increasing values of $\bar{\zeta}$ are found above $g_c$.
This indicates a modification of the spin configuration of the system upon increasing $g$,  related to
the development of a polarization.
On the other hand, we stress that an imbalanced occupation of the $+$ and $-$ bands alone is not sufficient to induce a finite polarization in the system. In fact,
this quantity is closely related to occupation of Rashba single-particle bands, and thus to the non-local spin structure
of the system. 

\begin{figure}[ht]
\vspace{0.2cm}
\centering
\includegraphics[scale=0.115]{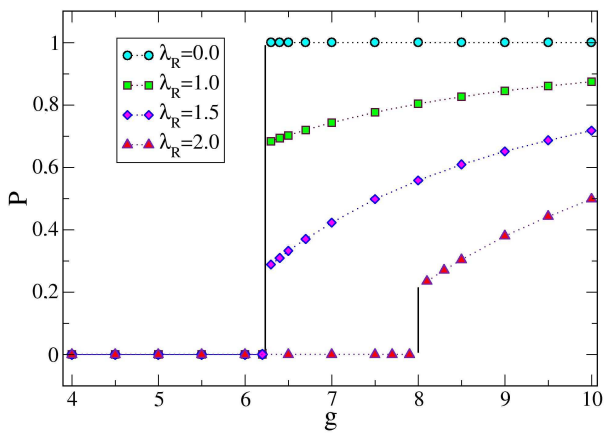}
\vspace{0.2cm}
\caption{(color online) Polarization of the optimal variational state, plotted in function of $g$ for different values of $\lambda_R$. 
Solid black lines indicate the phase transition (shifted in the case of $\lambda_R=2$). Units are specified in the text.}
\label{figura3}
\end{figure}

By analyzing the polarization $P$ as a function of $g$ and $\lambda_R$
we observe a first-order phase transition, analogous to that found in absence of SO, 
leading to finite polarization at strong repulsion (see Fig. \ref{figura3}).  
While $P$ is identically zero below $g_c$, a gradual polarization increase with respect to $g$, 
depending on $\lambda_R$, is found upon introduction of 
the SO coupling, beyond the critical point.
\newline
\newline
A slower convergence to the fully polarized state ($P=1$) is also observed for the larger $\lambda_R$, 
while a decrease of $P$ with respect to $\lambda_R$ is found at fixed $g$.
This suggests a clear tendency of the SO coupling  to effectively frustrate the spin alignment.
Moreover, large values of $\lambda_R$ can even cause a complete loss of
 polarization ($P=0$). This remarkable effect is visible in Fig.\ref{figura3} for $\lambda_R=2.0$, and can be
understood analytically through an expansion over $\zeta$ of the total energy $E_{{\rm VAR}}(n_+,\zeta)$
around $\zeta=0$.
Under the condition $\lambda_R>\sqrt{\pi n}/m$ only the $-$ band is occupied (see for instance Fig.\ref{figura2}, upper panel), and
one can prove that, to leading order in $\zeta$ 
\begin{equation}
E_{{\rm VAR}}(n_+=0,\zeta)\simeq C + \frac{n\zeta^2}{2m\lambda_R^2}-\frac{g}{4}\frac{n^2\zeta^2}{m^2\lambda_R^4} \,,
\end{equation}
where $C$  is constant with respect to $\zeta$.
Clearly, the Rashba coupling varies the relative importance of the two-body and one-body terms in the above equation,
changing the convexity of $E_{{\rm VAR}}$ with respect to $\zeta$.
Hence, the value of $\zeta$ that minimizes the energy becomes eventually zero for $\lambda_R > \sqrt{gn/(2m)}$.
Moreover, a numerical study of $E_{{\rm VAR}}(n_+=0,\zeta)$ reveals that $\zeta=0$ is a global
minimum in this context, and should represent the best approximation to the ground state, according to the variational principle.

\begin{figure}[ht]
\vspace{0.2cm}
\centering
\includegraphics[scale=0.115]{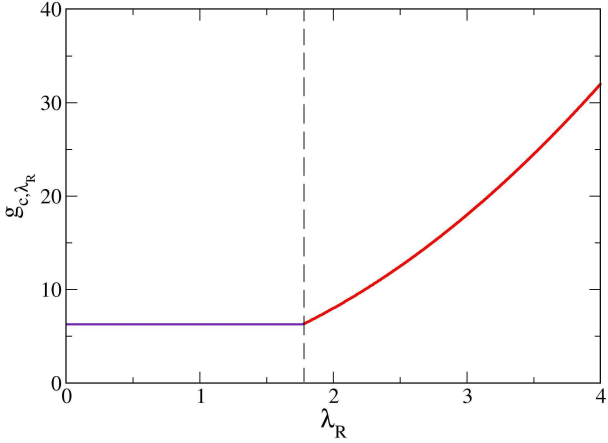}
\vspace{0.2cm}
\caption{(color online) Critical repulsive interaction coupling constant $g_{c,\lambda_R}$ as a function $\lambda_R$. 
The dashed vertical line separates the regimes of constant and variable $g_{c,\lambda_R}$. Units are specified in the text.}
\label{figura4}
\end{figure}

Since non-zero $P$ is only possible at finite $\zeta$, 
no spin polarization will be present in the system when $\lambda_R > \sqrt{gn/(2m)}$.
Equivalently, the critical value of $g$, beyond which finite polarization is developed, will depend on the SO coupling, and
can be expressed as
\begin{equation}
g_{c,\lambda_R}=
  \left\{
    \begin{array}{cc}
    2\pi/m   & \text{if } \lambda_R<\sqrt{\pi n}/m  \\
    2m\lambda_R^2/n    &\text{if } \lambda_R>\sqrt{\pi n}/m
  \end{array} \right.
\end{equation}
The critical $g$ will thus coincide with the value predicted in absence of SO coupling if $\lambda_R<\sqrt{\pi n}/m$,
but it will be shifted to higher values (stronger repulsion), for larger SO couplings (see Fig.\ref{figura4}).
As a whole, the Rashba coupling and the repulsive two-body coupling thus constitute complementary tools for an effective tuning
 of the system polarization, opening the way to the development of partially polarized states and control
of the phase transition.
\newline
\newline

As a final remark, we stress that the present approach reproduces, for $\lambda_R\rightarrow 0$, the mean field results presented
by Conduit \cite{cond2d} in absence of SO. On the other hand, the further inclusion of correlation terms was recently
shown \cite{cond2d,pilati} to produce a reduction of the critical coupling constant $g_c$, and a contextual smoothing of the
transition.
An analogous behavior is thus expected also in presence of the Rashba coupling, while the  SO
effects presented above should be roughly preserved. In fact SO couplings have a one-body structure, and their effects were recently
shown to be only marginally influenced by the correlation induced by the two-body repulsion \cite{vmc,dmc,chesi}.

\section{Conclusions}
The spin polarization of the two-dimensional Fermi gas in presence of contact repulsion and
Rashba SO coupling has been computed from a variational theory, based on the optimization
 of the wave function spinorial structure.
 As a result, we observe no polarization  below the critical repulsive coupling $g$ predicted in absence of SO coupling,
and confirm the presence of a phase transition.
Above criticity the polarization is determined by a competition between the repulsive interaction and the Rashba coupling.
While the two-body repulsion has a polarizing effect, the SO coupling tends to frustrate the spin alignment,
and can eventually lead to a complete loss of polarization. This mechanism determines a variability of the critical
$g$ as a function of $\lambda_R$, with a consequent shift of the phase transition. Contextually, partially polarized
states can be obtained by tuning $\lambda_R$.

\section{Acknowledgements}
We acknowledge  Francesco Pederiva and Enrico Lipparini for
useful discussions.

\begin{appendix}
\section{Analytical formulas}
We provide here an analytic computation of $E_{{\rm VAR}}(n_+,\zeta)$ defined in Eq. \eqref{evar}.
To make use of the convenient single-particle energy expressions given in Eq.\eqref{spen},
we first observe that $\hat{h}_{sp}=\hat{h}_{\zeta}-\hat{v}_{\zeta}$.
The variational energy can thus be recast into
\begin{align*}
E_{{\rm VAR}}(n_+,\zeta)=\,& E^{(0)}_{\zeta,n_+} + \langle\Phi^{\rm VAR}_{\zeta,n_+}|\hat{H}_{2B}|\Phi^{\rm VAR}_{\zeta,n_+}\rangle + \nonumber \\&
- \zeta \langle\Phi^{\rm VAR}_{h,n_+}|\hat{m}|\Phi^{\rm VAR}_{\zeta,n_+}\rangle \,.
\label{en1}
\end{align*}
The eigenvalue $E^{(0)}_{\zeta,n_+}$ defined in Eq. \eqref{e-egvl} can be computed as
 $E^{(0)}_{\zeta,n_+}=E^{(0),+}_{\zeta,n_+}+E^{(0),-}_{\zeta,n_+}$, where $E^{(0),\pm}_{\zeta,n_+}$
are defined as the sum of the single particle energies (see Eq. \eqref{spen}).

Regarding the $+$ band, the total single-particle energy contribution is given by the expression
\begin{align*}
E^{(0),+}_{\zeta,n_+}&=\frac{V}{(2\pi)^2}\int d^2\mathbf{k}\,\, \epsilon_{+}(\mathbf{k})\,\,  \Theta(\epsilon^F_{+}-\epsilon_{+}(\mathbf{k}))= \nonumber \\
&=\frac{\pi n_+^2}{m} + \frac{1}{2\pi}\frac{1}{3\lambda_R^2}\big[\left(\zeta^2+4\pi\lambda_R^2n_+ \right)^{3/2} -\zeta^3 \big]\nonumber
\end{align*}

Regarding the $-$ band, instead, one has to distinguish between the cases $\epsilon_-^F>-\zeta$ 
(single intersection with the $-$ band), and  $\epsilon_-^F<-\zeta$ (double intersection with the $-$ band):
\begin{align}
E^{(0),-}_{\zeta,n_+}=&\frac{V}{(2\pi)^2}\int d^2\mathbf{k}\,\, \epsilon_{-}(\mathbf{k})\,\,  \Theta(\epsilon^F_{-}-\epsilon_{-}(\mathbf{k}))= \nonumber \\
=&\frac{\pi n_-^2}{m} - \frac{1}{2\pi}\frac{1}{3\lambda_R^2}\big[\left(\zeta^2+4\pi\lambda_R^2n_- \right)^{3/2} -\zeta^3 \big] \nonumber \\
   &\qquad\qquad\qquad\qquad\qquad\qquad\qquad \text{if } \epsilon_-^F>-\zeta  \,, \hspace{3mm}  \nonumber  \\
=&\frac{n_-}{2\lambda_R^2}(\pi^2 n_-^2-\zeta^2+\lambda_R^2)-    \nonumber \\
&-\frac{1}{2\pi}\frac{1}{3\lambda_R^3} \big[\left(\lambda_R^2 + \pi n_- \right)^3-\left(\lambda_R^2 - \pi n_- \right)^3 \big] \nonumber \\
  &\qquad\qquad\qquad\qquad\qquad\qquad\qquad \text{if } \epsilon_-^F<-\zeta \, . \nonumber
\end{align}
The dependence on $n_-$ can be easily removed from the above equations through the relation $n_-=n-n_+$.

Analogously to $E^{(0),\pm}_{\zeta,n_+}$, the expectation value of the two-body interaction $\hat{H}_{2B}$
(indicated as $E^{(1)}$) and of the  operator $\hat{m}$ (indicated with $M$) are also defined as functions
of $n_{+}$ and $\zeta$. 

We provide here the analytical formulas for $E^{(1)}$ and $P$ distinguishing again between the two
 cases $\epsilon_-^F<-\zeta$ and $\epsilon_-^F>-\zeta$.

\begin{widetext}
$\epsilon_-^F>-\zeta$:
\begin{equation}
E^{(1)}=g\Big[ \frac{1}{4}(n_++n_-)^2-\left(\frac{\zeta}{4\pi \lambda_R^2}\right)^2 \left( \sqrt{\zeta^2+4\pi\lambda_R^2n_+}-\sqrt{\zeta^2+4\pi\lambda_R^2n_-} \right) \Big]
\end{equation}
\begin{equation}
M=\frac{\zeta}{2\pi\lambda_R^2}\left( \sqrt{\zeta^2+4\pi\lambda_R^2n_-}-\sqrt{\zeta^2+4\pi\lambda_R^2n_+} \right)
\end{equation}

$\epsilon_-^F<-\zeta$:
\begin{equation}
E^{(1)}=g\Big[ \frac{1}{4}(n_++n_-)^2-\left(\frac{\zeta}{4\pi \lambda_R^2}\right)^2 \left( \sqrt{\zeta^2+4\pi\lambda_R^2n_+}-\zeta-\frac{2\pi n_-}{m} \right) \Big]
\end{equation}
\begin{equation}
M=\frac{\zeta}{2\pi\lambda_R^2}\left(\zeta+\frac{2\pi n_-}{m} -\sqrt{\zeta^2+4\pi\lambda_R^2n_+} \right)
\end{equation}
\end{widetext}

Since in both cases $M$ contains $\zeta$  as an overall multiplicative factor, no polarization is  possible for $\zeta=0$, in line with the observations of Sec. II.
We remind, in this regard, that the polarization $P$ (see Eq.\eqref{P}) is related to $M$ by $P=M/n$ \,.
\end{appendix}

\end{document}